\documentclass[12pt]{iopart}
\usepackage{graphicx}
\usepackage{amssymb}
\usepackage{cite}
\begin{document}

\title[Creation of vibrationally-excited ultralong-range Rydberg molecules]
{Creation of vibrationally-excited ultralong-range Rydberg molecules in polarized and unpolarized cold gases of $^{87}$Sr}

%
%
%
%
%
%
%
%

\author{R. Ding, S. K. Kanungo, J. D. Whalen, T. C. Killian, F. B. Dunning}

\address{Department of Physics and Astronomy, Rice University, Houston, TX  77005-1892, USA}

\author{S. Yoshida, J. Burgd\"{o}rfer}

\address{Institute for Theoretical Physics, Vienna University of Technology, Vienna, Austria, EU}

\begin{abstract}
Photoexcitation rates for creation of ultralong-range Rydberg molecules
(ULRM) with 31$\lesssim n \lesssim41$ in both ground and excited vibrational levels in cold
($T\sim900$~nK) gases of polarized and unpolarized $^{87}$Sr are presented. The measured production
rates of the $\nu=0, 1$ and 2 vibrational levels reveal rather different
$n$ dependences which are analyzed by evaluating the Franck-Condon factors
associated with excitation of the different vibrational levels and molecular rotational states. In particular, for
gases of spin-polarized fermions, only Rydberg dimers with odd rotational
quantum numbers are excited due to the requirement that their wavefunctions be
anti-symmetric with respect to exchange. The data also demonstrate that
measurements of the formation of vibrationally-excited $\nu=1$ molecules can
furnish a probe of pair correlations over intermediate length scales extending from $\sim20$~nm to greater than 250~nm.
\end{abstract}

\maketitle

\section{Introduction}\label{S:into}
Ultralong-range Rydberg molecules (ULRM) have been the subject of much recent
interest because of their novel physical and chemical properties\cite{shaf18}.  In the
present work we focus on Rydberg dimer molecules which comprise a ground-state
atom weakly bound to a high-$n$ Rydberg atom by scattering of the Rydberg
electron.  While the existence of such molecules was first predicted
theoretically~\cite{gree00}, they have now been observed using a variety of
different atomic Rydberg states and a number of atomic species including
rubidium, cesium, and
strontium~\cite{bend09,li11,tall12,bell13,ande14,desa15,sass15,nied16,shaf18}. The
interaction between the excited Rydberg electron and ground state atom can be
approximated using a Fermi pseudopotential~\cite{ferm34}.
The resulting molecular potential can support a number of vibrational levels
which, for example, for principal quantum numbers $n\sim30$, have binding energies of
a few, to a few tens, of megahertz. Since the binding energies are so low,
Rydberg molecules can only be studied in cold molecular gases,
$T \lesssim 1$~mK. The binding energies decrease rapidly with
increasing $n$ scaling as $\sim 1/n^{6}$.

\begin{figure}[b]
\begin{center}
\includegraphics[width=10cm]{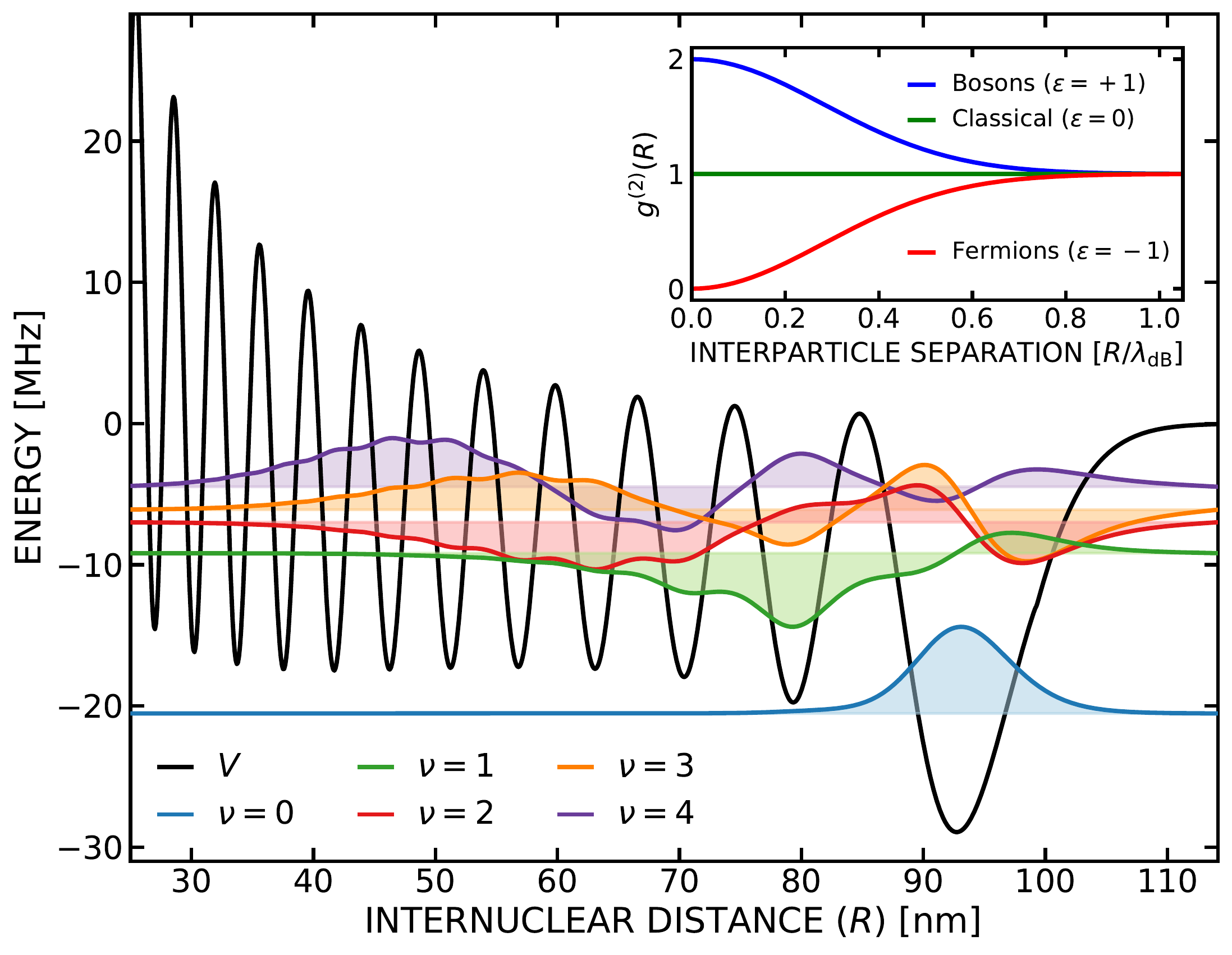}
\end{center}
\caption{\label{fig:mol potential}
Calculated molecular potential for a $5s34s$~$^{3}$S$_{1}$-$5s^{2}~^{1}S_{0}$
strontium atom pair (see Eq.\ref{eq:pseudopotential}) together with the calculated vibrational wavefunctions multiplied by the radial coordinate $R$ for
the $\nu$=0 to $\nu$=4 vibrational states.  The horizontal axis for
each wavefunction denotes its binding energy.  The inset shows the pair
correlation functions, $g^{(2)}(R)$, for cold thermal gases of non-interacting
identical bosons, fermions, and classical, i.e., distinguishable,
particles~\cite{nara99}.  The particle separations are expressed in units of
the thermal de Broglie wavelength $\lambda_{dB}=h/\sqrt{2\pi m k_{B}T}$
where $m$ is the atomic mass, $k_{B}$ the Boltzmann constant, and T the temperature.}
\end{figure}
An example of a molecular potential is depicted in
Fig.~\ref{fig:mol potential} together with the radial component
of the vibrational wavefunctions
for the $\nu$=0 to 4 vibrational levels.
The oscillatory structure in the molecular potential reflects the modulations
in the radial Rydberg electron probability density distribution.
The wavefunction for the ground $\nu$=0 vibrational state is strongly localized in the outermost well of the molecular potential which is located near the outer classical turning point of the Rydberg electron orbit.  The probability of photoexciting a $\nu=0$ dimer molecule therefore depends on the likelihood of finding a pair of ground-state atoms with the required initial internuclear separation, $R$.  Thus, by varying $n$, and hence the location of the potential minimum, measurements of dimer formation can be used to probe the spatial dependence of the pair correlation function, $g^{(2)}(R)$, in a cold gas~\cite{nara99}.  This has been exploited in earlier work  to probe non-local pair correlations in cold gases of (bosonic) $^{84}$Sr and (fermionic) $^{87}$Sr over length scales of $\sim50-150$~nm using Rydberg molecules with $31\leq n\leq 45$~\cite{whal19a,whal19}.  These studies clearly demonstrated the effects of quantum statistics, i.e., of bunching in a thermal gas of  $^{84}$Sr and antibunching due to Pauli exclusion in a spin-polarized gas of  $^{87}$Sr.

As seen in Fig.~\ref{fig:mol potential}, the wavefunctions for higher excited vibrational states $\nu =1, 2, \cdots$
extend
to smaller internuclear separations  than for the $\nu$=0 states which
suggests that measurements of the formation rates for the different
vibrationally-excited dimer levels might be used to probe spatial correlations over a
broader range of $R$.  This we examine in the present work where we
compare results for the formation of the different molecular vibrational
states  in a cold ($T\sim 900 $nK) gas of spin-polarized $^{87}$Sr with
results for an unpolarized $^{87}$Sr sample.  $^{87}$Sr atoms have a sizable
nuclear spin, $I=9/2$, resulting in a total angular momentum $F=9/2$ for the
$5s^{2}~^{1}S_{0}$ ground state and a large number of magnetic sublevels,
$m_{F} =-9/2, -7/2 . . .7/2, 9/2$.  Because of this ten-fold degeneracy, a
statistical population of ground-state $^{87}$Sr atoms provides a good
approximation to a gas of uncorrelated, i.e., classical, particles.  (The bosonic isotopes of strontium have $I=0$ and thus no degeneracy in the ground state.  Here, when we refer in general to bosonic isotopes, we assume they are spin polarized.)

The experimental results are interpreted through comparison to calculated
rates for molecule formation that incorporate an effective  Franck-Condon factor and
account for pair correlations.
The results presented here further demonstrate that studies of Rydberg molecule formation can provide a valuable probe of spatial correlations in quantum gases over a sizable (and previously inaccessible) range of internuclear separations together with a test of the present theoretical understanding of the molecular potentials and wavefunctions involved.

\section{Experimental approach}\label{experimental approach}
  As is apparent from the inset in Fig.~\ref{fig:mol potential}, for cold gases of non-interacting identical bosons or fermions, quantum statistics only begin to have a significant effect on
 $g^{(2)}(R)$ at small interparticle spacings $R\lesssim0.8~\lambda_{dB}$, where $\lambda_{dB}$ is the thermal de Broglie wavelength, and their effects only become readily apparent at even smaller interparticle spacings, say $R\sim 0.4~\lambda_{dB}$.   Earlier work has shown that an $^{87}$Sr gas can be readily cooled to temperatures of $\sim800-900$~nK corresponding to $\lambda_{dB}\sim200$~nm enabling study of the effects of quantum statistics at interparticle spacings $\lesssim160$~nm.  An interparticle spacing of 160~nm corresponds to the radius, $R_{n}$, of $5sns~^{3}S_{1}$ Rydberg atoms (given by $R_{n}\sim2(n-\delta)^{2} a_{0}$ where $\delta\sim3.37$ is the quantum defect and $a_{0}$ the Bohr radius) with $n\sim40$.  Smaller interparticle separations can be investigated through formation of molecules with smaller values of $n$.  However, it is difficult to extend measurements to values of $n\lesssim30$ because production of a Rydberg molecule requires that the ground-state atom density, $\rho$, in the trap be such that there is a significant likelihood of finding ground-state atom pairs with the necessary spacing which, for $n\sim30$, necessitates cold atom densities $\rho\gtrsim3\times 10^{13}$~cm$^{-3}$.  In addition, measurements for $n\gtrsim45$ are challenging as the spacings between neighboring vibrational levels, which decrease rapidly as $n$ increases, become very small making them difficult to resolve with our existing laser linewidth of $\sim$300~kHz.

To more directly compare molecule formation in polarized and unpolarized
gases it is advantageous to produce samples with very similar temperatures and
density distributions.  This is challenging because achieving sub-$\mu$K
temperatures requires evaporative cooling. For an unpolarized sample, energy
transfer during collisions allows the sample to continuously thermalize as
the trap depth is lowered.  In contrast, for a spin-polarized sample there
are no $s$-wave collisions at these low temperatures and therefore
thermalization is suppressed.  To overcome this problem a mixture of $^{84}$Sr and $^{87}$Sr is trapped and sympathetic cooling used to obtain the desired temperature.

\begin{figure}[hb]
\begin{center}
\includegraphics[width=10cm]{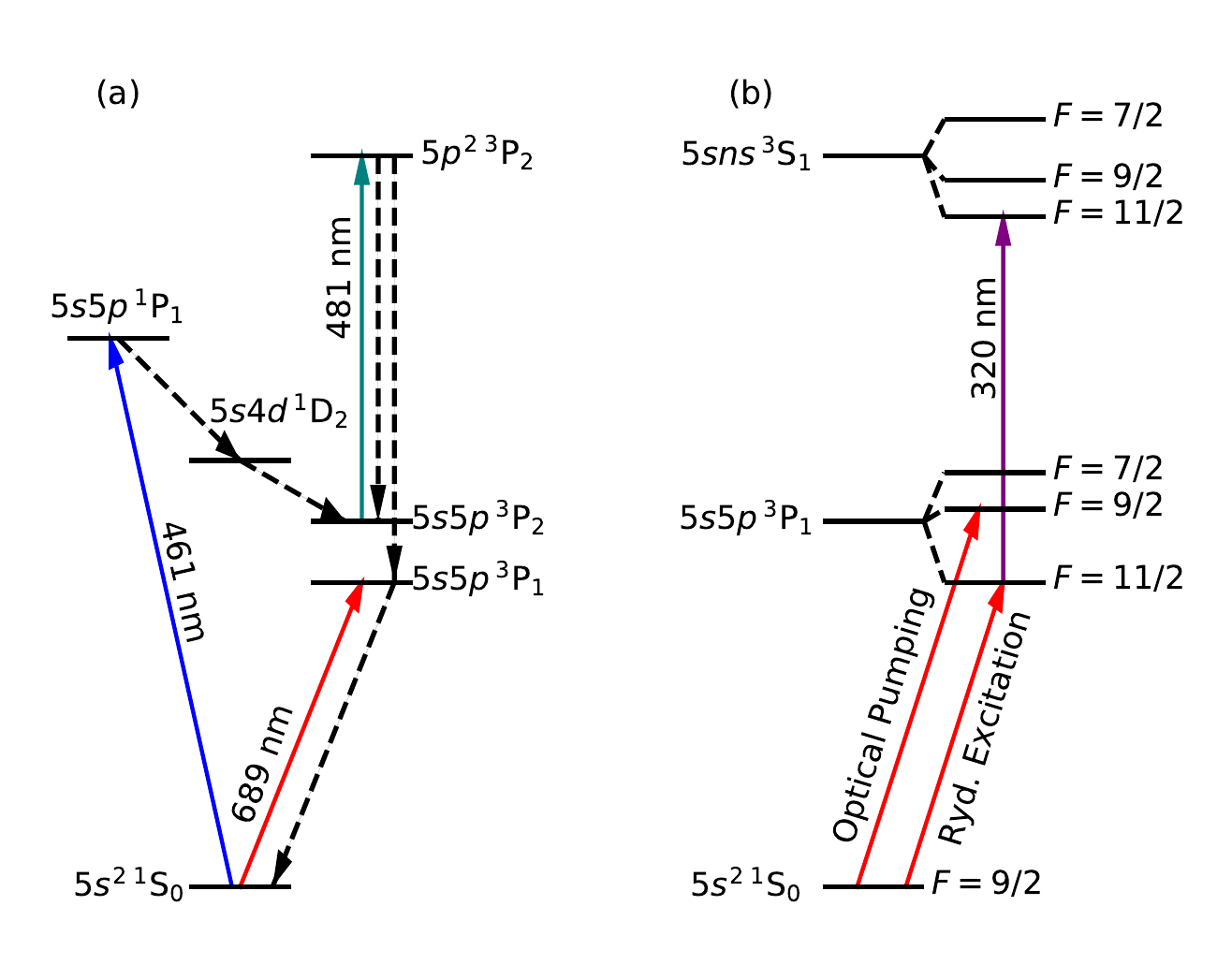}
\end{center}
\caption{\label{fig:term diagram}
 a) Schematic partial term diagram for strontium showing the levels used for laser cooling and repumping. The dashed lines represent the spontaneous decay paths involved in magnetic trapping and in repumping.  b) Schematic diagram of the relevant transitions used for optical pumping and for two-photon excitation to 5sns~$^{3}$S$_{1}$ Rydberg states including the hyperfine structure.}
\end{figure}
\begin{figure}[hb]
\begin{center}
\includegraphics[width=10cm]{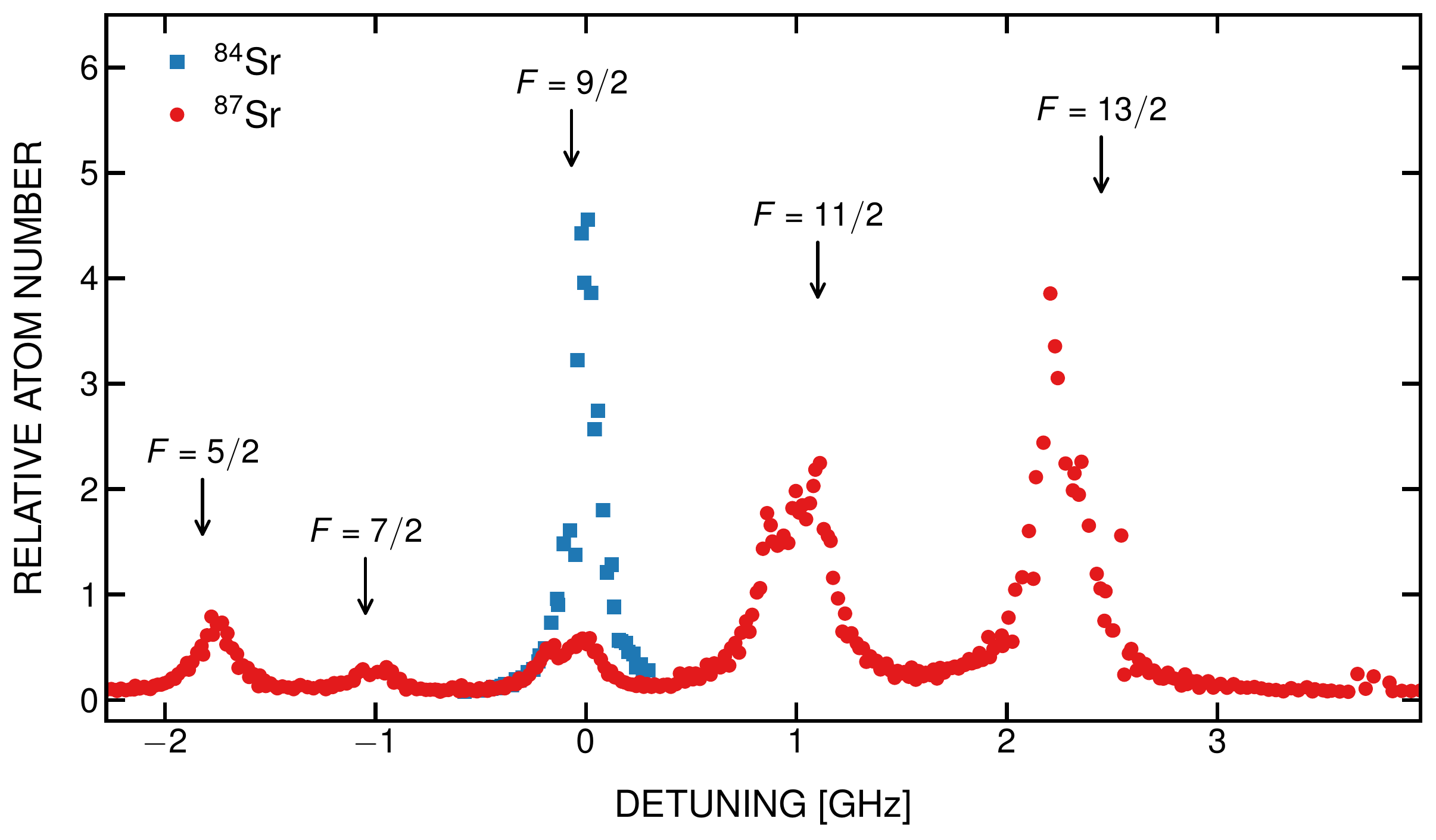}
\end{center}
\caption{\label{fig:atom signal}
Ground-state $^{87}$Sr atom signal observed as the 481~nm repump laser is scanned over the 5s5p~$^{3}$P$_{2}$-5p$^{2}~^{3}$P$_{2}$ transition.  The initial 5s5p ~$^{3}$P$_{2}$ hyperfine level associated with each $^{87}$Sr feature is indicated (see text).  Data recorded using $^{84}$Sr are also included. }
\end{figure}

The present cooling protocol~\cite{dees09,stel14}
can be understood with reference to
Fig.~\ref{fig:term diagram}
which presents a partial term diagram for strontium.  Strontium atoms emerging from a Zeeman slower are first cooled to temperatures of a few millikelvin in a magneto-optical trap (MOT) operating on the 461~nm $5s^{2}~^{1}S_{0}-5s5p~^{1}P_{1}$ transition.  Atoms in the excited state, however, have a small probability of decaying into the long-lived $5s5p~^{3}P_{2}$ metastable state via the $5s4d~^{1}D_{2}$ state.  Those $^{3}P_{2}$ atoms formed in weak-field-seeking states become trapped in the MOT magnetic field which therefore serves as a magnetic trap \cite{nage03}.  Atoms are allowed to accumulate in this trap to build up high atom densities.  Because of isotope shifts in the $^{1}S_{0}-^{1}P_{1}$ transition, $^{87}$Sr and $^{84}$Sr atoms are loaded sequentially into the magnetic trap to allow the 461~nm laser to be separately tuned for each isotope.  (Sequential loading also allows the relative populations of each isotope to be varied by controlling the loading times.)  After loading the magnetic trap the atoms are returned to the $^{1}S_{0}$ ground state via the $5s5p~^{3}P_{1}$ state using a repump laser operating on the $5s5p~^{3}P_{2}-5p^{2}~^{3}P_{2}$ transition at 481~nm.  However, as illustrated in
Fig.~\ref{fig:atom signal},
 which shows the ground-state atom signal observed as the repump laser is scanned, the presence of hyperfine structure in $^{87}$Sr results in multiple spectral features which complicates the repump process.  (The observed splittings indicate that the structure is associated primarily with the hyperfine splitting of the lower $5s5p~^{3}$P$_{2}$ state.)  Efficient and simultaneous repumping of all the $^{87}$Sr hyperfine states, as well as admixed $^{84}$Sr atoms, therefore requires the presence of multiple laser frequencies which are generated by broadening the laser spectrum by modulating its drive current and superposing sidebands using an EOM.  Following repumping, both isotopes are simultaneously cooled to $\sim2\mu$K using a MOT operating on the 689~nm $5s^{2}~^{1}S_{0}\rightarrow5s5p~^{3}P_{1}$ intercombination line.  This is accomplished using three separate laser frequencies: one laser is tuned to the $^{1}S_{0}\rightarrow^{3}P_{1}$ transition in $^{84}$Sr, the other two to the $^{1}S_{0}~F=9/2\rightarrow^{3}P_{1}~F=11/2$ and $^{1}S_{0}~F=9/2\rightarrow^{3}P_{1}~F=9/2$ transitions in $^{87}$Sr. The atoms are then loaded into a ``pancake-shaped" optical dipole trap (ODT) formed using a 1.06~$\mu$m laser beam in the form of a flat sheet with a width of $\sim260\mu$m and thickness $\sim26\mu$m in the center of which is a ``dimple" of $\sim60\mu$m diameter created using a second laser beam incident near normal to the plane of the sheet.

Spin-polarized samples are obtained by optically pumping the atoms in the
ODT.  A magnetic bias field of 7.6~G is established which produces a Zeeman
splitting of $\sim650$~kHz between adjacent magnetic sublevels and a series
of $\sigma^{+}$-polarized laser pulses tuned to the $5s^{2}~^{1}S_{0}~F=9/2$
to $5s5p~^{1}P_{1}~F=9/2$ transition is applied to transfer the population to
the $M_{F}=+9/2$ magnetic sublevel.  Once optical pumping is complete the
magnetic field is reduced to 1~G to preserve a quantization axis.
Detailed spectroscopic measurements\cite{whal19,whal19a} showed that optical pumping transfers $>90\%$ of the ground-state atoms to the $m_{F}=+9/2$ magnetic sublevel.
Experiments with unpolarized samples are undertaken in zero magnetic field.

 The atoms in the ODT are cooled to $\sim900$~nK through evaporative cooling.  Given the same initial ratio of $^{84}$Sr to $^{87}$Sr in the ODT, the final temperature of a sample of spin-polarized $^{87}$Sr atoms will always be higher than that for an unpolarized sample because of heating due to photon scattering during optical pumping.  To obtain polarized and unpolarized samples with similar densities and temperatures the ratio of $^{84}$Sr and $^{87}$Sr atoms loaded into the ODT is varied - the greater the fraction of $^{84}$Sr the colder the final sample.  Once evaporation is complete, a light pulse resonant with the $5s^{2}~^{1}S_{0}\rightarrow5s5p~^{3}P_{1}$ transition in $^{84}$Sr is applied to remove these atoms from the trap through light scattering.  (The $^{84}$Sr-$^{87}$Sr isotope shift is sufficient that the laser pulse does not lead to any detectable heating of the $^{87}$Sr atoms.)  The final atom number and temperature is determined by releasing the atoms from the trap and, after a fall time of $\sim7$~ms, measuring the spatial extent of the cloud using absorption imaging on the $5s^{2}~^{1}S_{0}\rightarrow5s5p~^{1}P_{1}$ transition.

 Following preparation of an $^{87}$Sr sample, Rydberg excitation spectra are recorded using pulsed two-photon excitation.  The first (689-nm) photon is $\sigma^{+}$ polarized (and is blue detuned $\sim14.8$~MHz from the transition to the $5s5p~^{3}P_{1}~F=11/2$ level) and the second (319-nm) photon is $\pi$ polarized and tuned to excite final $5sns~^{3}S_{1}~F=11/2$ Rydberg states.  The ODT is turned off during the excitation pulses to eliminate AC Stark shifts.  The product Rydberg atoms/molecules are detected through field ionization by applying voltage pulses to electrodes that surround the trap.  The product electrons are directed towards, and detected by, a dual microchannel plate (MCP) detector whose output is fed to a multichannel scalar (MCS).  The number of Rydberg atoms/molecules created in a single excitation pulse is kept small to avoid both saturating the detector and blockade effects, and data are accumulated following many laser pulses to build up good statistics.

\section{Theoretical method}\label{S:Theoretical method}

The interaction between the quasi-free Rydberg electron and the neutral
ground state atom is very weak and dominated by the short-ranged
Fermi pseudopotential~\cite{ferm34}.
The effective potential of an ULRM is therefore approximately given by
\begin{eqnarray}
\label{eq:pseudopotential}
V(\vec{R})&=&2\pi
\frac{\hbar^{2}}{m_{e}}a_{s}\vert\psi(\vec{R})\vert^{2}+6\pi\frac{\hbar^{2}}{m_{e}}a_{p}^{3}\vert\vec{\nabla}
\psi(\vec{R})\vert^{2},
\end{eqnarray}
where $\psi(\vec{R})$ is the electronic wavefunction, $m_{e}$ the electron
mass, $e$ the electronic charge,
and $a_{s}$ and $a_{p}$ are the $s$- and $p$-wave scattering lengths.
In the following we consider Rydberg dimers formed by $5sns$~$^3S_1$ Rydberg
states. For such a spherically symmetric charge cloud
the molecular potential is isotropic and depends only on the
internuclear distance $R$ between the Rydberg core and the ground state atom.
Therefore, the eigenstates of Rydberg dimers can be written as~\cite{thom18}
\begin{equation}
\Psi_{\nu,\Lambda,M_\Lambda}(R,\theta,\phi)
= {\cal R}_{\nu, \Lambda}(R)
Y_{\Lambda}^{M_{\Lambda}}(\theta,\phi)
\end{equation}
with $\nu$ the vibrational quantum number,
$\Lambda$ the rotational quantum number,
and $M_\Lambda$ the projection of $\vec{\Lambda}$ onto the magnetic field axis.
(For Rydberg $S$-states the third Euler angle becomes cyclic and the
Wigner rotation matrix reduces to a spherical harmonic
$Y_{\Lambda}^{M_{\Lambda}}$.)

The rate of excitation of a Rydberg molecule is governed by experimental factors such as laser intensity and sample density as well as by the
atomic dipole transition strength and a Franck-Condon-type factor
for characterizing the overlap between the initial unbound ground-state
atom-pair wavefunction and the ULRM wavefunction. Since the
interaction between the Rydberg electron and the ground state atom
is typically very weak, the molecular potential
[Eq.~(\ref{eq:pseudopotential})] is evaluated in first-order perturbation
theory, i.e., the unperturbed electronic wavefunction of the
Rydberg atom is used in Eq.~(\ref{eq:pseudopotential}).
Moreover, the electronic transition strength depends only on the atomic
Rydberg wavefunction, in particular on the principal quantum number $n$ and
quantum defect $\delta$ of the Rydberg atom,
$\langle d \rangle^2 \sim (n-\delta)^{-3}a^{2}_{0}$,
but is independent of the molecular level to be formed.
The Franck-Condon factor governing the production rate from a particular initial two-body scattering state $\chi_{\vec{k}}^{\pm}(\vec{R})$ is given by
\begin{eqnarray}
\label{eq:Franck-Condon}
{\cal F}^{\pm}_{\nu,\Lambda}(\vec{k})
=\int d^{3}R {\cal R}_{\nu,\Lambda}(R) Y_{\Lambda}^{M_{\Lambda}}(\theta,\phi)
   \chi_{\vec{k}}^{\pm* }(\vec{R})
\end{eqnarray}
where  $\vec{R}$ is the relative coordinate, $\pm$ is the parity, and
$\hbar \vec{k}$ is the relative momentum of the two neighboring ground state atoms that eventually form
the Rydberg dimer through photoexcitation of one of the atoms to
a Rydberg state.
Assuming that the interaction between the two ground-state atoms is negligibly small and that the potential of the ODT is constant over the length scale of the Rydberg atom, the properly symmeterized initial two-body scattering states are given by
\begin{equation}
\label{eq:init_scat}
\chi^{\pm}_{\vec{k}}(\vec{R}) = \frac{1}{\sqrt{2}}
\left( e^{ i \vec{k} \cdot \vec{R}} \pm  e^{ -i \vec{k} \cdot \vec{R}} \right)
\, .
\end{equation}
For ground-state $^{87}$Sr atoms, if the gas is spin polarized, the scattering state must have odd parity.  Otherwise, we describe the ensemble as an admixture of scattering states with both parities.

By assuming a Lorentzian profile, we define
the spectral excitation density for a given molecular
state $(\nu, \Lambda)$, fixed relative wave vector $\vec{k}$,
and parity $\pm$ as
\begin{equation}
\label{eq:spectrum_nu}
f^{\pm}_{\nu,\Lambda}(\vec{k},\omega) = \frac{1}{\pi}
  |{\cal F}^{\pm}_{\nu,\Lambda} (\vec{k})|^2
  \frac{ \Gamma/2}{(\hbar \omega +k^2/(2 \mu)-E_{\nu,\Lambda})^2 + (\Gamma/2)^2}
\end{equation}
where $E_{\nu,\Lambda}$ is the binding energy of the Rydberg molecule,
$\omega$ is the laser detuning from the resonant excitation of $5sns$ $^3S_1$
Rydberg atoms, and $\mu$ is the reduced mass which is half the $^{87}$Sr mass.
In the current setting, the experimental resolution determines
the effective width $\Gamma$ ($\sim 300$~kHz) which is much larger than
the lifetime broadening of the Rydberg molecule.
The total spectral excitation density from all states of parity $\pm$ follows then from Eq.~(\ref{eq:spectrum_nu}) as the sum over
all molecular states $(\Lambda, \nu)$ and the
average over the thermal distribution at given temperature $T$ over
the relative momenta of the atom pairs.  Assuming the system is far from quantum degeneracy, this yields
\begin{eqnarray}
\label{eq:spectrum}
f^{\pm}(\omega)
= \sum_{\nu,\Lambda} (2 \Lambda + 1)
  \left( \frac{1}{2\pi \mu k_{B}T} \right)^{3/2}
  \int d^3 k e^{-k^2/(2 \mu k_B T)} f^{\pm}_{\nu,\Lambda}(\vec{k},\omega) \, .
\end{eqnarray}
The factor, $2 \Lambda + 1$, is due to the degeneracy of the
$M_{\Lambda}$ levels in the Rydberg molecule for a given $\nu$ and $\Lambda$.
Typically, the energy shifts associated with rotational excitation are small
(for $n=30$ and $\nu=0$ the rotational constant is $\sim 20$~kHz )
and individual  rotational levels cannot be resolved.
However, the vibrational levels ($\nu = 0,1,2$) have significantly
larger energy spacing ($\sim 30$~MHz at $n=30$ and
$\sim 4$~MHz at $n=40$) and can be resolved. In such cases, an excitation
strength for each vibrational level can be separately
determined by summing over the
thermally-averaged Franck-Condon factors
$\langle |{\cal F}^{\pm}_{\nu,\Lambda}|^2 \rangle$
from all rotational levels or approximated
by integrating $f^{\pm}(\omega)$ over a frequency window centered
at a given vibrational level
\begin{equation}
\label{eq:ex_strength}
P^{\pm}_{\nu}
= \sum_\Lambda  (2\Lambda+1)
\langle |{\cal F}^{\pm}_{\nu,\Lambda}|^2 \rangle
\simeq \int_{E_{\nu,\Lambda}/\hbar-\Delta}^{E_{\nu,\Lambda}/\hbar+\Delta}
f^{\pm}(\omega) d\omega \, .
\end{equation}
where $\Delta$ is much larger than $\Gamma$ but is small compared to the
vibrational level spacing, i.e., of the order of 1~MHz.
When the radial wavefunction of a vibrational state
is well localized at $R = R_n$,
the excitation strength for states of $\pm$ parity can be approximated as
\begin{equation}
\label{eq:ex_strength0}
P^{\pm}_\nu \simeq 4 \pi g_{\pm}^{(2)}(R_{n})
  \left|
    \int dR \, R^2 {\cal R}_{\nu,\Lambda}(R)
   \right|^2
\end{equation}
Since, for $30 \le n \le 41$, the centrifugal potential only
adds nearly a constant energy shift to the molecular potential
and the resulting molecular wavefunctions ${\cal R}_{\nu,\Lambda}(R)$
are nearly independent of $\Lambda$.
In Eq. (\ref{eq:ex_strength0}), we have introduced
\begin{eqnarray}
g_{\pm}^{(2)}(R) &=& \frac{1}{4\pi} \sum_\Lambda  (2\Lambda+1)
\left( \frac{1}{2\pi \mu k_{B}T} \right)^{3/2}
  \int d^3 k e^{-k^2/(2 \mu k_B T)}
\nonumber \\
&& \times
  \left|
    \int d\cos\theta d\phi \, Y_{\Lambda}^{M_{\Lambda}}(\theta,\phi)
   \chi_{\vec{k}}^{\pm*} (\vec{R})
  \right|^2
\, .
\label{eq:g2_ini}
\end{eqnarray}

The plane waves (Eq.~\ref{eq:init_scat}) appearing in Eq.~(\ref{eq:g2_ini}),
can be expanded in partial waves with well-defined
exchange symmetry (or parity),
\begin{equation}
\chi^{\pm}_{\vec{k}}(\vec{R}) = 4 \sqrt{2} \pi
 \sum_{\Lambda=0}^\infty \sum_{M_\Lambda=-\Lambda}^\Lambda
  i^\Lambda {\cal P}^{\pm}_\Lambda Y_{\Lambda}^{M_\Lambda, *}(\theta_k,\phi_k)
  \chi_{k,\Lambda,M_\Lambda} (\vec{R})
\end{equation}
with $\theta_k, \phi_k$ the polar angles of $\vec{k}$,
\begin{equation}
\chi_{k,\Lambda,M_\Lambda} (\vec{R}) = j_\Lambda(k R)
 Y_{\Lambda}^{M_\Lambda}(\theta,\phi)
\label{eq:sph_wave}
\end{equation}
($j_\Lambda(k R)$ : spherical Bessel function) and
\begin{equation}
{\cal P}^{\pm}_\Lambda = \frac{1}{2} \left( 1 \pm (-1)^\Lambda \right)
\end{equation}
restricts the wavefunction to even angular momenta $\Lambda$ for symmetric
$(+)$ and odd $\Lambda$ for anti-symmetric $(-)$ two-body scattering states.  This yields
\begin{eqnarray}
g_{\pm}^{(2)}(R)
&=& 8 \pi \sum_\Lambda (2\Lambda+1) {\cal P}^{\pm}_\Lambda
  \left( \frac{1}{2\pi \mu k_{B}T} \right)^{3/2}
\nonumber \\
&& \times
  \int d k \, k^2 e^{-k^2/(2 \mu k_B T)} |j_\Lambda(k R)|^2 = 1\pm e^{-2\pi R^{2}/ \lambda^{2}_{dB}}\, .
\label{eq:g2_int}
\end{eqnarray}
For a spin-polarized gas of $^{87}$Sr atoms, the excitation strength for a transition to a single localized vibrational state $\nu$ is
$P_\nu^{\rm pol} =
{\cal Q}_{\rm pol}^+ P_\nu^{+} + {\cal Q}_{\rm pol}^- P_\nu^{-} = P^{-}_{\nu}$ defining ${\cal Q}_{\rm pol}^+ = 0$ and ${\cal Q}_{\rm pol}^- =1$.  For unpolarized fermions in which all the $M_{F}$ levels are populated with equal probability the excitation strength is $P_\nu^{\rm unpol} = {\cal Q}_{\rm unpol}^+ P_\nu^{+} + {\cal Q}_{\rm unpol}^- P_\nu^{-}$ where
\begin{equation}
{\cal Q}^{\pm}_{\rm unpol} = \frac{1}{2F+1}\left\{
  \begin{array}{ll}
    F & \mbox{for $+$} \\
    F + 1  & \mbox{for $-$}
  \end{array}
\right. \, .
\end{equation}
We can thus express the excitation strength for both polarized and unpolarized gases for excitation to a vibrational state well localized at $R=R_{n}$ as
\begin{equation}
P^{pol/unpol}_{\nu}\simeq4\pi g^{(2)}(R_{n})\left\vert\int dR~R^{2}\mathcal{R}_{\nu,\Lambda}(R)\right\vert^{2}
\label{eq:strength}
\end{equation}
where
\begin{equation}
g^{(2)}(R)={\cal Q}^{-} g^{(2)}_{-} (R) + {\cal Q}^{+}g^{(2)}_{+}(R) = 1-\epsilon e^{-2\pi R^{2}/ \lambda^{2}_{dB}}
\label{eq:cor function}
\end{equation}
is the appropriate correlation function for the sample, assuming weak interactions and thermal equilibrium far from quantum degeneracy.  Here $\epsilon = {\cal Q}^{-} - {\cal Q}^{+}$.  $g^{(2)}(R)$ takes the following forms:  for spin polarized fermions
\begin{equation}
\label{eq:g2_pol}
g^{(2)}(R) = g_-^{(2)}(R)
= 1 - e^{-2\pi R^{2}/\lambda_{dB}^{2}} \, ,
\end{equation}
for an unpolarized ensemble with $F=9/2$
\begin{equation}
\label{eq:g2_unpol}
g^{(2)}(R) = g_{\rm unpol}^{(2)}(R)
= 1 -0.1 e^{-2\pi R^{2}/\lambda_{dB}^{2}} \, .
\end{equation}
and for a gas of spin polarized bosons
\begin{equation}
\label{eq:g2_boson}
g^{(2)}(R) = g_+^{(2)}(R)
= 1 + e^{-2\pi R^{2}/\lambda_{dB}^{2}} \, .
\end{equation}
Equation \ref{eq:strength} is accurate for transitions to the molecular ground state ($\nu=0$) because this state is typically localized in the outer well of the potential at $R=R_{n}$ (Fig.~\ref{fig:mol potential})\cite{whal19}.  Thus, measurements of
$P_{\nu=0}$ can be used to extract information on the correlation function.  Since the radial integrals in  Eq.~(\ref{eq:strength}) are common
for spin-polarized and unpolarized gases ,
the ratio $\xi_{\nu=0} = P_{\nu=0}^{\rm pol}/P_{\nu=0}^{\rm unpol}$
can be related to the pair correlation function for spin-polarized
gases~\cite{whal19}.  If it is assumed that $g^{(2)}(R)$ for the unpolarized gas is given by Eq~\ref{eq:g2_unpol}, then $g^{(2)}(R)$ for the polarized gas
\begin{equation}
g^{(2)}(R_{n})= \xi_{\nu=0}(1-0.1 e^{-2\pi R_{n}/\lambda^{2}_{dB}})
\label{eq:unpol _gas}
\end{equation}
and $\xi_{\nu=0}$ can be found from the ratio of the experimental signal rates for transitions in polarized and unpolarized gases after taking into account different Clebsch-Gordan coefficients and any variations in experimental parameters such as laser intensities and sample densities and temperature\cite{whal19}

\begin{figure}[t]
\begin{center}
\includegraphics[width=8cm]{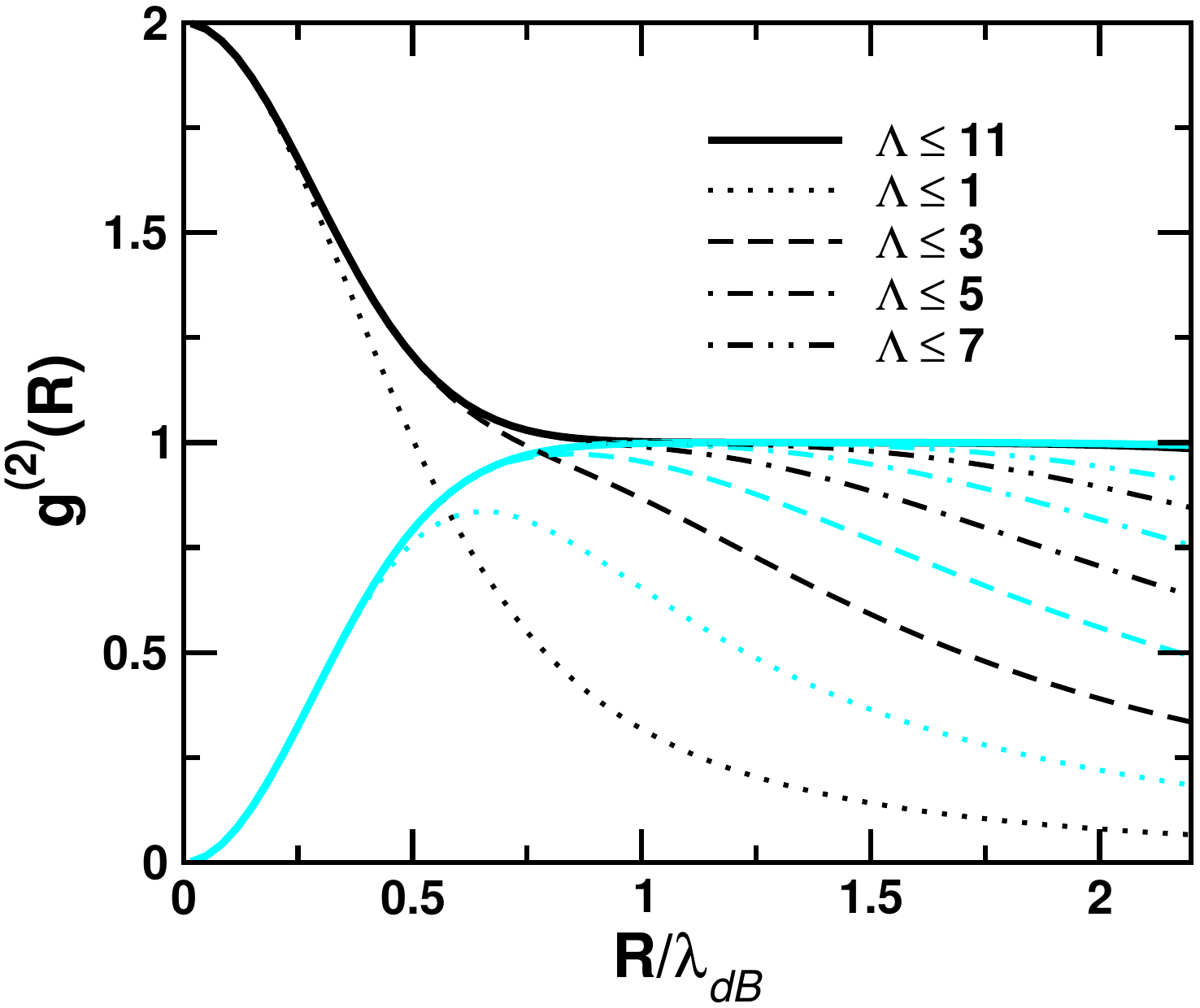}
\end{center}
\caption{\label{fig:g2_L}
$g^{(2)}(R)$ for $^{84}$Sr (bosons, in black) and spin-polarized$^{87}$Sr
(fermions, in light blue/gray) as function of $R/\lambda_{dB}$
calculated with rotational states included up to the values of $\Lambda$ indicated.  For bosons
(fermions) only states with even (odd) values of $\Lambda$ contribute to
$g^{(2)}(R)$ (see text).
}
\end{figure}

Equation~(\ref{eq:g2_int}) indicates that, depending on the temperature $T$
of the atomic ensemble, a significant number of rotational levels $\Lambda$
contribute to the observed pair correlation function.
Figure~\ref{fig:g2_L} shows the relative contributions to
$g_{+}^{(2)}(R)$ associated with states with different values of $\Lambda$
as a function of $R/\lambda_{dB}$, for identical bosons.
As expected, the maximum in
$g_{+}^{(2)}(R)$ at small values of $R/\lambda_{dB}$ is
associated primarily with $\Lambda=0$ states, i.e., $s$-waves.
As $R/\lambda_{dB}$ increases, due,
for example, to an increase in sample temperature, higher-$\Lambda$ states become
accessible and become increasingly important while the $\Lambda = 0$
contribution is reduced. Indeed, for $R/\lambda_{dB} \sim 1$,
$\Lambda=2$ states, i.e., the $d$-wave, becomes the dominant
contribution to $g_+^{(2)}(R)$ which has by then become close to its limiting
value $g_+^{(2)}(R)=1$.  Further increase in $R/\lambda_{dB}$ results in little
change of $g_+^{(2)}(R)$, although the relative contributions from
higher-$\Lambda$ states steadily grow.  For spin-polarized fermions,
the $s$-wave contribution
is excluded because of anti-symmetry and $g_-^{(2)}(R)$ vanishes
for $R/\lambda_{dB} \to 0$.  As $R/\lambda_{dB}$ increases, $p$-wave
and successively higher odd-order partial waves become accessible and
$g_-^{(2)}(R)$ approaches its limiting value of one.

\begin{figure}[th]
\begin{center}
\includegraphics[width=7cm]{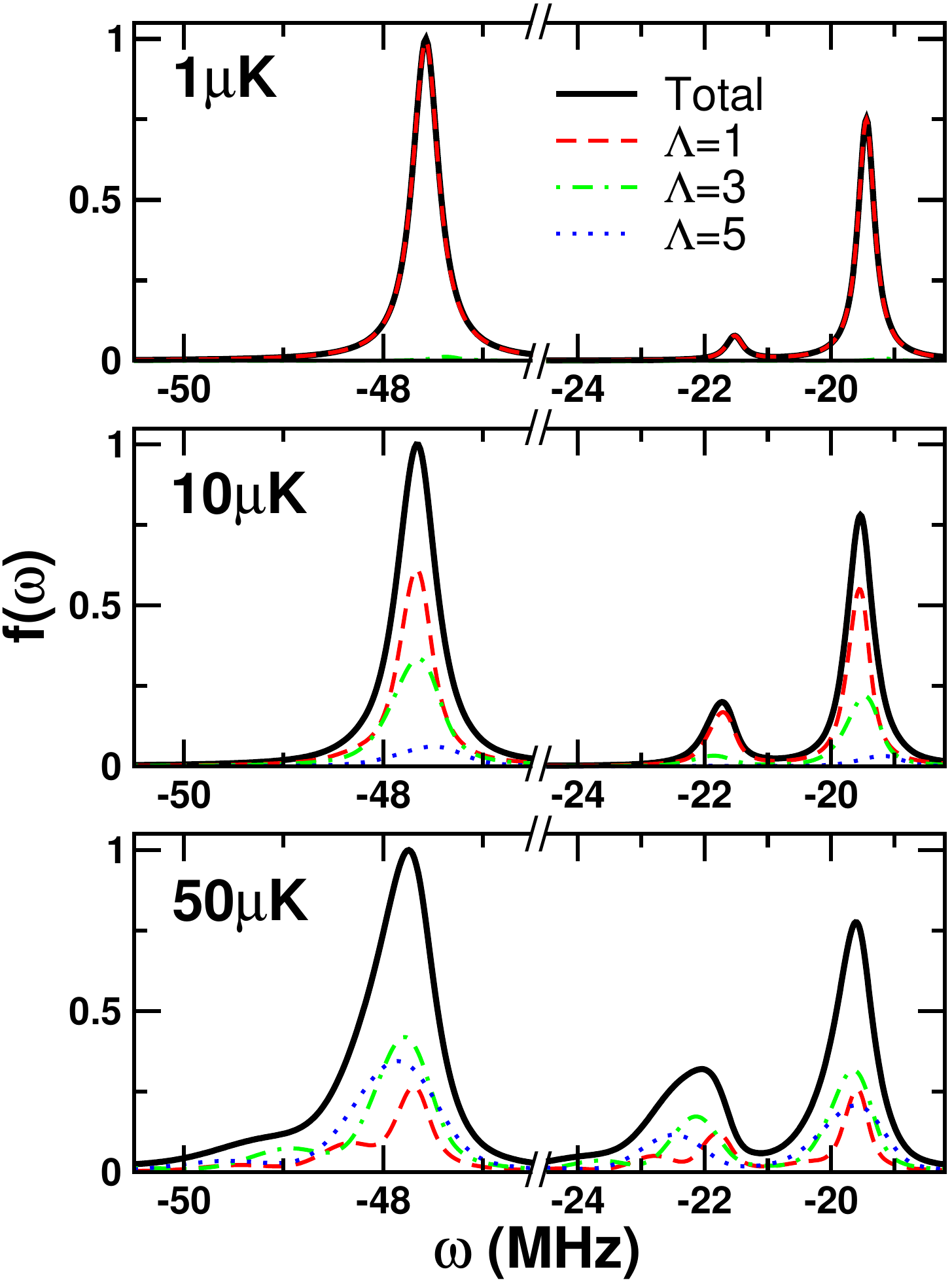}
\end{center}\caption{\label{fig:temp}
Excitation spectra (Eq.~\ref{eq:spectrum})
for ULRMs (black solid lines) associated with the
$5s30s$ $^3S_1$ Rydberg state calculated for a spin-polarized
$^{87}$Sr gas at the various temperatures indicated. The contributions from
each rotational level are also displayed
(red dashed line : $\Lambda=1$, green dot-dashed line : $\Lambda=3$,
and blue dotted line : $\Lambda = 5$).
}
\end{figure}

The contributions from
various $\Lambda$ levels to the calculated excitation
spectra (Eq.~\ref{eq:spectrum_nu}) are shown in Fig.~\ref{fig:temp}.
For spin-polarized fermions
the excitation spectrum is dominated by the $\Lambda = 1$ rotational state
at low temperature (1~$\mu$K). As temperature increases, the contributions
from higher excited rotational levels become non-negligible.
The peak positions of $f^-(\omega)$ for different $\Lambda$ nearly coincide
since the spacing between different rotational levels are smaller than the
thermal line broadening. Furthermore, since $\Lambda$ is preserved during the
Franck-Condon transition, the contribution to rotational energy splitting
from the centrifugal potential present in both the initial and final states
largely cancels out.  The contributions of different rotational channels to the Rydberg molecule excitation spectrum were recently discussed in reference \cite{sous19}.

\begin{figure}[th]
\begin{center}
\includegraphics[width=10cm,bb=50 200 550 760]{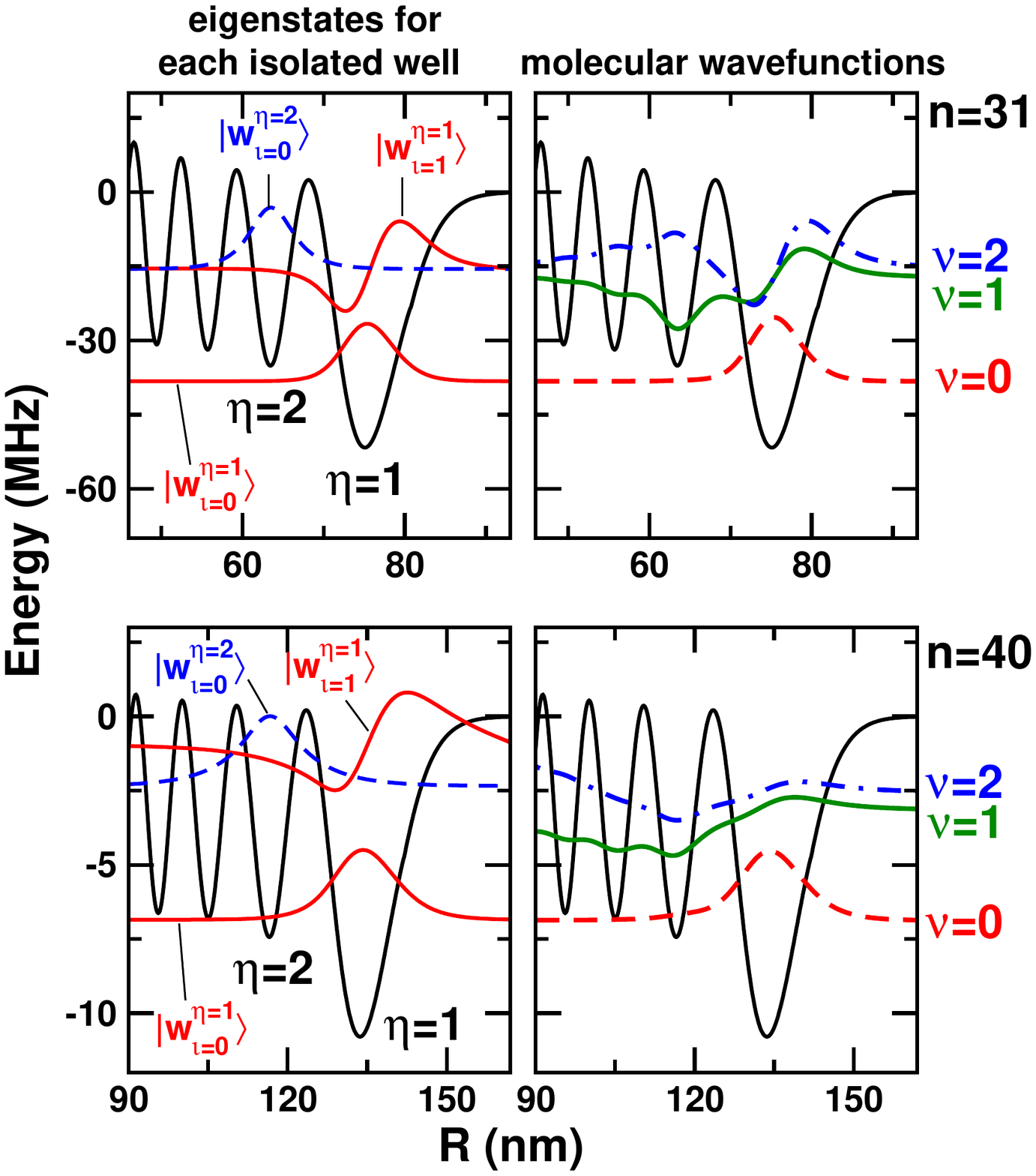}
\end{center}
\caption{\label{fig:wells}
Molecular potentials and associated eigenwavefunctions
for $n=31$ (upper frames) and 40 (lower frames) states with $\Lambda = 0$.
In the right column the molecular wavefunctions for $\nu=0$
(dashed lines in red), $\nu =1$ (solid line in green), and
$\nu = 2$ (dot-dashed line in blue) states are plotted.
In the left column the wavefunctions
$|w^{\eta=1}_{\iota=0,1}\rangle$ of the outer-most well
and $|w^{\eta=2}_{\iota=0}\rangle$
of the next nearest well are shown (see text). Solid lines (red) are those for the outermost well
and the dashed lines (blue) those for the neighboring well.
The wave functions are multiplied by the radial coordinate $R$ and
their base lines are shifted by their eigenenergies.
}
\end{figure}

Unlike the case for ground-state molecules, the molecular wavefunctions for excited Rydberg dimers span multiple wells (see Figs.~\ref {fig:mol potential}, \ref{fig:wells}).  We therefore explore the degree of localization of each ULRM molecular wavefunction in a specific well and how this affects the ability to use excitation spectra for excited dimers to extract information on $g^{(2)}(R)$.  To this end we construct a set of pseudostates $|w_{\iota}^\eta\rangle$
$(\iota = 0,1, \cdots)$ that are eigenstates of each isolated potential
well $\eta$ $(\eta=1,2,3, \cdots$ with $\eta = 1$ the outermost well)
thereby removing the influence of the adjacent potential wells. The molecular
wavefunction can then be (approximately) viewed as a
coherent superposition of eigenstates $|w^\eta_\iota\rangle$ associated
with each isolated well. For example, the lowest energy level
$|w^{\eta=1}_{\iota=0}\rangle$ of the outermost well ($\eta=1$)
lies below the minimum of the inner potential wells $(\eta=2,3,\cdots)$
and closely approximates the true ground vibrational state $\nu=0$
of the Rydberg dimer which is well localized in the outermost well near
$R_{n,\eta=1} \simeq 1.87 (n-\delta)^2 a_{0}$ and is undistorted by the
presence of adjacent wells (see Fig.~\ref{fig:wells}).
For the excited vibrational states the molecular wavefunctions are less
localized (see Figs.~\ref{fig:mol potential} and \ref{fig:wells})
and the extraction of the pair correlation function becomes more complicated.
For example, at $n=30$ the first excited state $|w^{\eta=1}_{\iota=1}\rangle$
of the outermost well is nearly degenerate with the lowest energy state
$|w^{\eta=2}_{\iota=0}\rangle$ of the second well.
Therefore, the molecular wave function for $\nu = 1$ can be approximated
by a coherent superposition of two single-well eigenstates
$c_{\eta=1}|w^{\eta=1}_{\iota=1}\rangle \mp
c_{\eta=2}|w^{\eta=2}_{\iota=0}\rangle$.

For the $\nu = 1$ wavefunction,
whose delocalized probability distribution spans two adjacent
wells, the thermally averaged Franck-Condon factor may be written
\begin{eqnarray}
\langle |{\cal F}_{\nu,\Lambda}|^2 \rangle &\propto&
  \int d k \, k^2 e^{-k^2/(2 \mu k_B T)}
   \left|
    c_{\eta=1} j_\Lambda(k R_{n,\eta=1})
    \int dR \, R^2 w^{\eta=1}_{\iota=1}(R)
\right.
\nonumber \\
&& \left.
    \mp c_{\eta=2} j_\Lambda(k R_{n,\eta=2}) \
    \int dR \, R^2 w^{\eta=2}_{\iota=0}(R)
   \right|^2 \,
\end{eqnarray}
assuming that (for small $T$) the spherical Bessel functions for the $k$-values that contribute to the transitions are  nearly constant within a single well centered
at $R =  R_{n,\eta}$. Because of the node
in the wavefunction $w^{\eta=1}_{\iota=1}(R)$ located in the
$\eta=1$ well, the overlap integral
$\int dR \, R^2 w^{\eta=1}_{\iota=1}(R) $ is typically small, and thus
the Franck-Condon factor can be simplified to
\begin{eqnarray}
\langle |{\cal F}_{\nu,\Lambda}|^2 \rangle \propto
  \int d k \, k^2 e^{-k^2/(2 \mu k_B T)} |j_\Lambda(k R_{n,\eta=2})|^2
   \left|
     c_{\eta=2} \int dR \, R^2 w^{\eta=2}_{\iota=0}(R)
   \right|^2 \,
\label{eq:FC_nu1}
\end{eqnarray}
with $R_{n,\eta=2} = 1.6 (n-\delta)^2 a_{0}$.
Summing over contributions from all $\Lambda$ levels with appropriate weighting for the polarization state of the gas,
the excitation strength becomes approximately
\begin{equation}
\label{eq:ex_strength1}
P_{\nu=1} \simeq 4 \pi g^{(2)}(R_{n,\eta=2})
  \left|
    c_{\eta=2} \int dR \, R^2 w^{\eta=2}_{\iota=0}(R)
   \right|^2 \, .
\end{equation}
With increasing $n$, the energy of $|w^{\eta=2}_{\iota=0}\rangle$ becomes
smaller than $|w^{\eta=1}_{\iota=1}\rangle$ (see Fig.~\ref{fig:wells})
and, correspondingly, the $\nu=1$ molecular state becomes
increasingly dominated by the $|w^{\eta=2}_{\iota=0}\rangle$ state.
For Franck-Condon factors with well-localized transition points
[Eq.~(\ref{eq:FC_nu1})]
the ratio $\xi_{\nu = 1} = P_{\nu = 1}^{\rm pol}/P_{\nu = 1}^{\rm unpol}$
can be used to probe the correlation function but over a range of $R$
different from that for the $\nu = 0$ states.
However, as contributions from other wells $(\eta > 2)$ become increasingly
important (for example, for higher $n$) the ratio
$\xi_{\nu=1}$
no longer  probes the pair correlation locally but provides an average
of $g_-^{(2)}(R)$ over a range of $R$ weighted by $|c_{\eta}|^2$.
For even higher vibrational states (for example $\nu=2$), the contributions from
inner wells ($\eta > 2$) can no longer be neglected.
As the molecular wavefunction becomes increasingly delocalized  reliable  extraction of the pair correlation function becomes difficult.

\section{Results and Discussion}
\label{Results and Discussion}
\begin{figure}[b]
\begin{center}
\includegraphics[width=10cm]{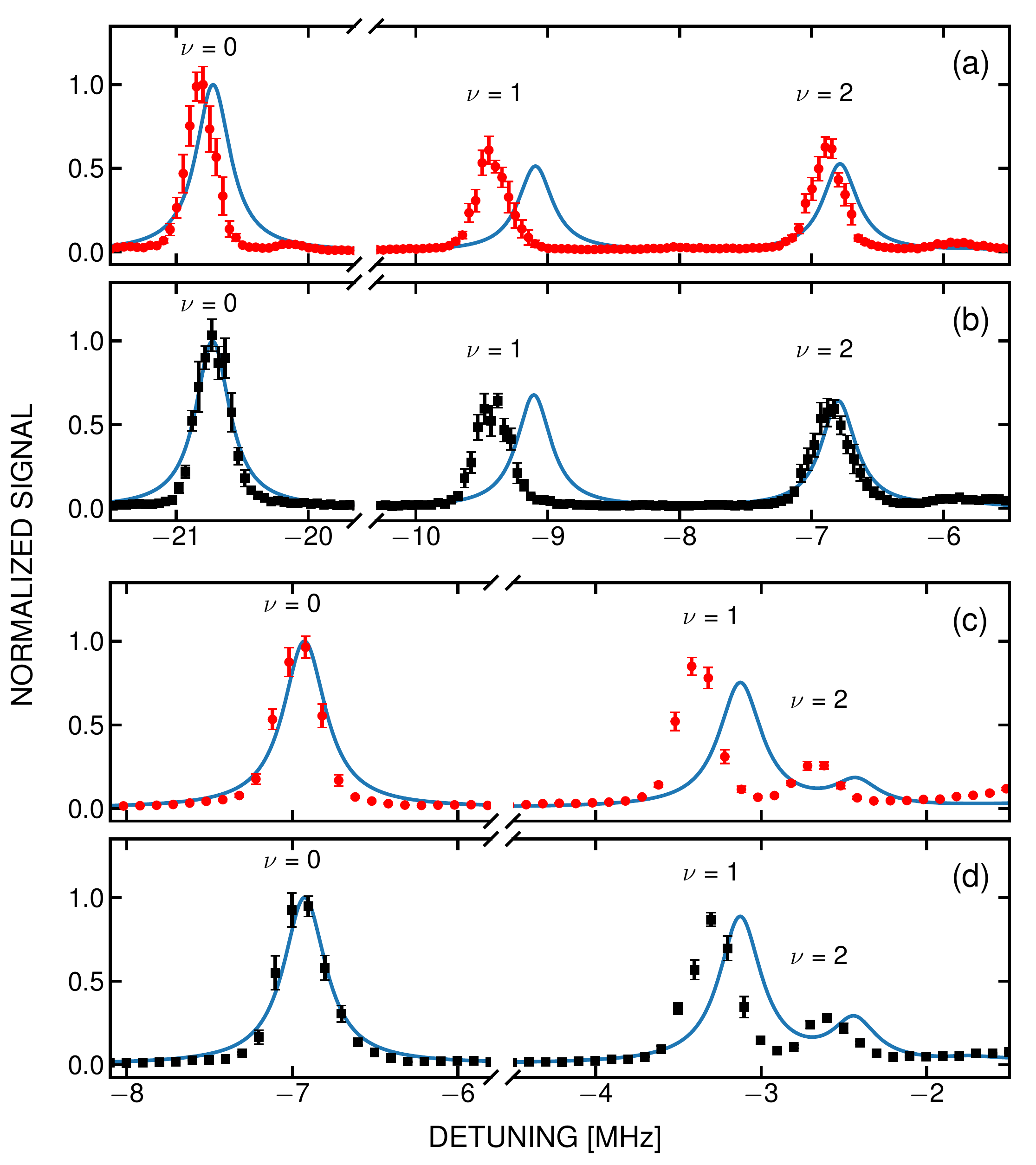}
\end{center}
\caption{\label{fig:spectra}
Experimental Rydberg excitation spectra recorded using (a,c) polarized and (b,d)
unpolarized  $^{87}$Sr cold gases with $T\sim900$~nK. (a,b) are the spectra for
5s34s~$^{3}$S$_{01}$-5s$^{2}~^{1}$S$_{0}$ molecules and (c,d) are
for 5s40s~$^{3}$S$_{1}$-5s$^{2}~^{1}$S$_{0}$ molecules.
The spectra are normalized such that the peaks of the $\nu$=0 features are of
equal height.  The solid lines show the predicted excitation spectra (see
text).  The calculated spectra have been convolved with a Lorentzian of
300~kHz width.}
\end{figure}
Experimental Rydberg excitation spectra recorded at $n=34$ and 40 using both polarized and unpolarized cold, $T\sim900$~nK, $^{87}$Sr gases are shown in Fig.~\ref{fig:spectra}.
The spectra are normalized such that the peaks associated with the formation
of $\nu$=0 ground-state Rydberg dimers are of equal height.
The actual sizes of the
excitation features seen in different experimental runs depend on many factors
including the intensities of the excitation lasers, the trapped atom density, the laser detuning, the dipole matrix elements, the Clebsch-Gordan coefficients, and the excitation strengths $P_{\nu}$ [Eqs.~\ref{eq:ex_strength} and \ref{eq:strength}].  Within a single panel in Fig~\ref{fig:spectra} all factors other than $P_{\nu}$ are the same, and the relative heights are proportional to $P_{\nu}$,
 thereby providing a more direct
test of the calculated effective
Franck-Condon factors and their underlying dependence on the pair correlation
function.  Multiple features are present in each spectrum that result from creation of different dimer vibrational states.
Figure~\ref{fig:spectra} also includes the predictions of calculations
[Eq.~(\ref{eq:spectrum})] using
the Fermi pseudopotential [Eq.~(\ref{eq:pseudopotential})]
with the effective $s$- and $p$-wave scattering lengths,
$a_{s}(k=0)=-13.3a_{0}$ and $a_{p}=9.7a_{0}$.
The positions and relative sizes of the observed features are in good general
agreement with the theoretical predictions.  However, as seen in Fig.~\ref{fig:spectra} the relative sizes of the $\nu=1$ and $\nu=2$
features are strongly $n$-dependent.  For $n=34$ the relative sizes
of the $\nu=1$ and $\nu=2$ features are comparable, whereas for
$n=40$ the $\nu=1$ feature is dominant.

\begin{figure}[b]
\begin{center}
\includegraphics[width=10cm]{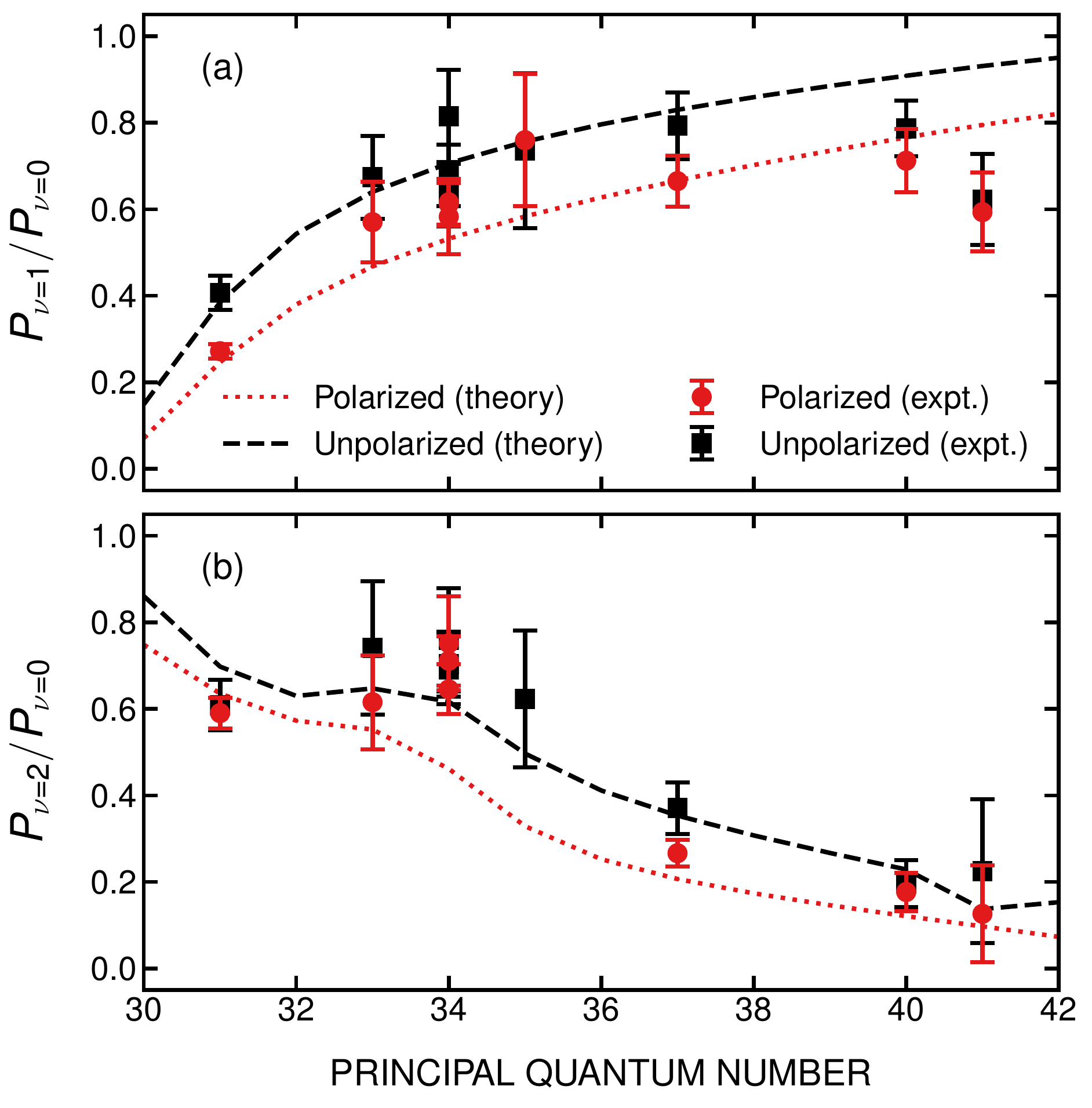}
\end{center}
\caption{\label{fig:n-dependence}
Measured ($\circ$) and calculated (- - -) $n$ dependence of the
production rates for a) $\nu$=1 and b) $\nu$=2 states relative to
that for $\nu$=0 states in polarized and unpolarized $T\sim 900nK ~^{87}$Sr samples.
{\bf }
}
\end{figure}
  Figure~\ref{fig:n-dependence} shows the integrated experimental signals for transitions to the $\nu=1$ (top) and $\nu=2$ states (bottom), normalized by the integrated signals for the $\nu=0$ states, for various $n$.  These ratios should equal the theoretically-calculated ratios of excitation strengths $P_{\nu=1,2}/P_{\nu=0}$.  The integrated experimental signals were obtained by fitting the different features with a pseudo-Voigt profile and determining the area under the resulting curve.  Interestingly, the relative excitation strengths of the $\nu=1$ and
$\nu=2$ features display very different $n$-dependences.  The relative
strength of the $\nu=1$ feature decreases markedly
with decreasing $n$, whereas that of the $\nu=2$ feature
increases substantially,  behavior that is well reproduced by theory.

 For $\nu = 1$ states, the calculated ratio of the excitation strengths for $\nu=1$ and $\nu=0$ states can be written using
Eqs.~\ref{eq:strength} and \ref{eq:ex_strength1} as
\begin{equation}
\frac{P_{\nu=1}}{P_{\nu=0}}
=|c_{\eta=2}|^2  \frac{g^{(2)}(R_{n,\eta=2})}{g^{(2)}(R_{n,\eta=1})}
  \frac{\left|
     \int dR \, R^2 w^{\eta=2}_{\iota=0}(R)
   \right|^2}
  {\left|
     \int dR \, R^2 w^{\eta=1}_{\iota=0}(R)
   \right|^2} \, .
\end{equation}
The $w^{\eta=1,2}_{\iota=0}$
states represent the ground states of nearly harmonic potential wells
pointing to similar $n$-dependences of the integrals for both $\eta=1$ and 2.
Therefore, since $R_{n,\eta=1}$ and $R_{n,\eta=2}$ are similar, this results in a ratio of the $g^{(2)}(R)$ that remains close to unity for the present range of $n$.
The strong $n$-dependence seen in the $\nu=1$ to $\nu=0$ production ratios must thus be associated principally with the weights $|c_{\eta=2}|^2$.
Furthermore, the calculated ratios $P_{\nu=1}/P_{\nu=0}$
for polarized samples are, on average, somewhat smaller than those
for unpolarized samples. This is due to the fact that $R_{n,\eta=2}$ is slightly less than $R_{n,\eta=1}$ and thus the ratio
$g^{(2)}(R_{n,\eta=2})/g^{(2)}(R_{n,\eta=1})$ for a polarized sample is slightly less than one (see the inset in Fig.~\ref{fig:mol potential})
leading to the small decrease  in $P_{\nu=1}/P_{\nu=0}$.
As discussed in the previous section
(see Fig.~\ref{fig:wells}), as $n$ increases the molecular state becomes increasingly dominated by
the $|w^{\eta=2}_{\iota=0}\rangle$ contribution and
$|c_{\eta=2}|^2$ increases. Therefore, the observed $n$-dependence
mirrors the
dominance of the $|w^{\eta=2}_{\iota=0}\rangle$ contribution to the $\nu=1$ vibrational state. For $\nu=2$, however,
the contributions from other wells ($\eta > 2$) become more significant
and the peaks in the molecular wavefunctions
shift towards smaller values of $R$, relative to the size of the atom, with increasing $n$.
This is reflected in the observed $n$-dependence of the $\nu=2$ features.

As demonstrated in earlier work that focused on the $\nu=0$ state~\cite{whal19}, pair correlation functions can be obtained from measurements of the ratio, $\xi_{\nu=0}= P_{\nu=0}^{pol}/P_{\nu=0}^{unpol}$, which can be determined from the relative molecular production rates in polarized and unpolarized samples.  Ideally such measurements should be undertaken using identical samples with the exception that one is polarized, the other unpolarized.  While, for the measurements reported here we attempt to match the sample conditions as closely as possible, differences remain.  The ratio of the measured production rates must be corrected for small differences in the intensities of the photoexcitation lasers, in laser detunings, in the sample temperatures and densities and density distributions, as well as for the differences in the two-photon electronic transition matrix elements, i.e., Clebsch-Gordan coefficients, when creating Rydberg molecules in polarized and unpolarized gases.

\begin{figure}[t]
\begin{center}
\includegraphics[width=10cm]{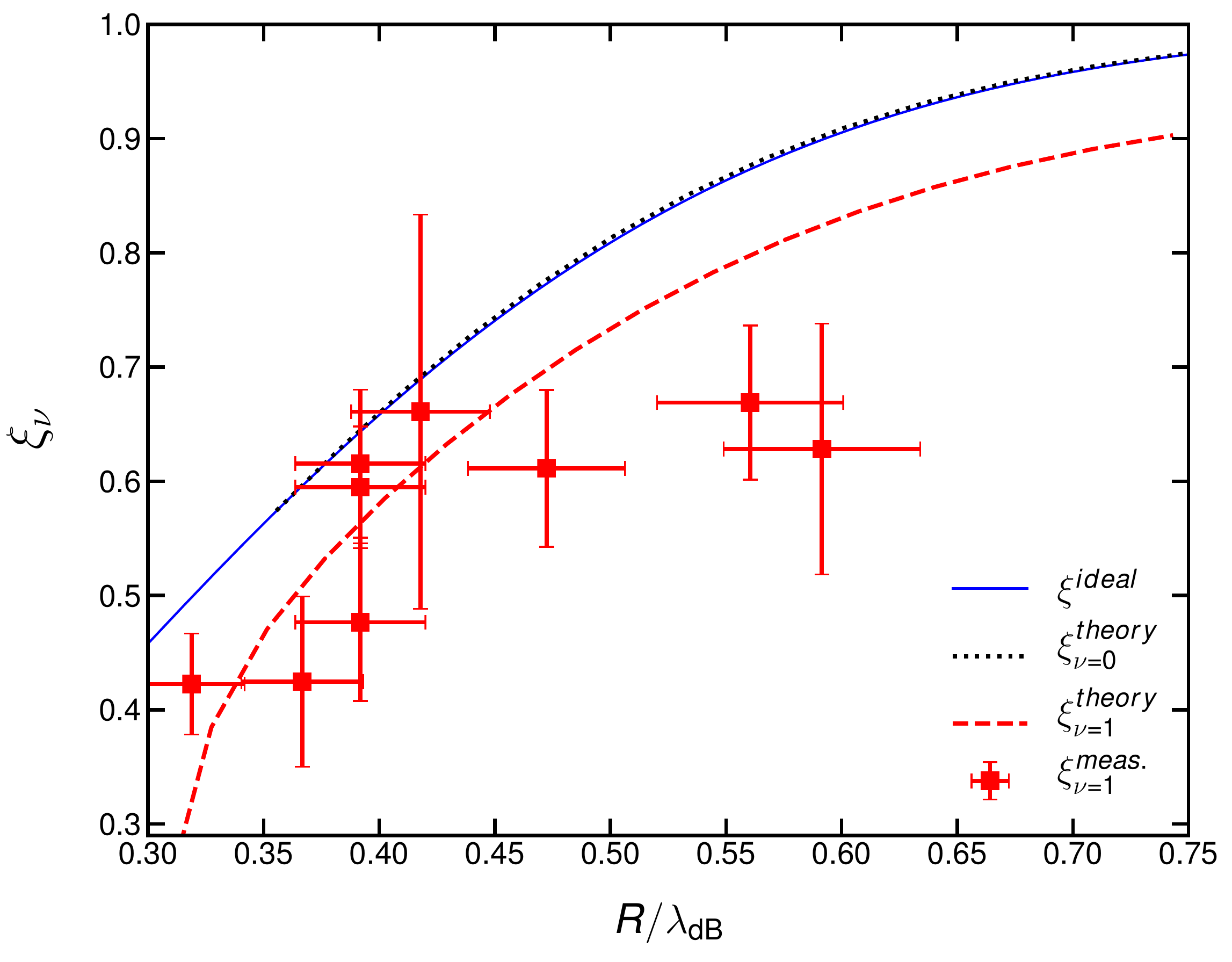}
\end{center}
\caption{\label{fig:g2}
Ratios, $\xi_{\nu}$, of the ULRM excitation strengths in polarized and unpolarized samples of $^{87}$Sr as a function of $R/\lambda_{dB}$, with $R=1.87(n-\delta)^{2}a_{0}^{2}$ for $\nu=0$
states and $R=1.6(n-\delta)^{2}a_{0}^{2}$ for $\nu=1$ states.
The figure includes the results of measurements, $\xi_{\nu=1}^{meas}$, for the $\nu=1$ state together with theoretical predictions for the $\nu=0$ and 1 states, $\xi_{\nu=0}^{theory}$ and $\xi_{\nu=1}^{theory}$, and the ``ideal'' ratio $\xi^{ideal}=g_{-}^{(2)}(R)/g_{unpol}^{(2)}(R)$ (see text).
{\bf }
}
\end{figure}

Figure~\ref{fig:g2} shows the similarly determined ratios of excitation strengths, $\xi_{\nu=1}^{meas}= P_{\nu=1}^{pol,meas}/P_{\nu=1}^{unpol,meas}$, for the $\nu=1$ level in polarized and unpolarized samples.  Also included in Fig~\ref{fig:g2} are the theoretically-predicted ratios of the excitation strengths, $\xi_{\nu=0}^{theory}=P_{\nu=0}^{pol,theory}/P_{\nu=0}^{unpol,theory}$ and $\xi_{\nu=1}^{theory}=P_{\nu=1}^{pol, theory}/P_{\nu=1}^{unpol,theory}$,
for the $\nu=0$ and $\nu=1$ states, respectively.  For reference, Fig.~\ref{fig:g2} also shows the ratio to be expected under the simple zeroth-order ``ideal'' assumption that  $\xi^{ideal}=g_{-}^{(2)}(R)/g_{unpol}^{(2)}(R)$.
As noted in earlier work\cite{whal19}, $\xi_{\nu=0}^{theory}$ closely matches $\xi^{ideal}$.  The predicted values of $\xi_{\nu=1}^{theory}$ are somewhat smaller than $\xi_{\nu=0}^{theory}$, which results because, while the contributions from potential wells other than $\eta=2$ are small, they are significant.   Nonetheless, the predicted values are in reasonable agreement with experiment, although the discrepancy seen at the largest values of $R/\lambda_{dB}$, i.e., the largest values of $n$, remains to be explained.
 However, the pronounced decrease in $\xi_{\nu=1}^{meas}$ at the smaller values of $R$ provides clear evidence of the effects of antibunching, and the data demonstrate that (for $31\leq n\leq41$) measurements of the $\nu=1$ vibrational state can provide a probe of pair correlation functions at values of $R$ that are somewhat smaller than can be realized using $\nu=0$ states and where the effects of quantum statistics become increasingly important.

\section{Conclusions}
\label{conclusions}
Measurements of the photoexcitation of ultralong-range Rydberg molecules
(ULRM), specifically Rydberg dimers,
can provide an {\it in~situ}
probe of pair correlations in an ultracold gas that,
with an appropriate choice of $n$, can be tuned over previously inaccessible
length scales that extend from $\sim20$ to greater than 250~nm.  (Quantum gas microscopes can resolve
correlations on length scales on the order of half the wavelength of
light~\cite{mazu17,bakr10} whereas
inelastic loss from spin flips and three-body recombination probe two- and
three-body spatial correlations at shorter length scales~\cite{burt97}.)
The present approach provides a valuable new window into intermediate-range
phenomena such as the formation of Halo states~\cite{koeh06,jens04}
and Efimov trimers~\cite{chin10}, and allows study of correlation functions for scattering states involving atom pairs with large s-wave scattering lengths.
Furthermore, since the time scale for molecule formation is short,
$\sim1\mu$s, the present approach is suitable for {\it in~situ}
probing of non-equilibrium
dynamics in quantum gases.

\ack
The authors thank R. G. Hulet for the loan of equipment.  Research supported
by the AFOSR (FA9550-14-1-0007), the NSF (1600059), the Robert A. Welch
Foundation (C-0734 and C-1844), the FWF (Austria)(FWF-SFB041 ViCom, and
FWF-W1243).  The Vienna scientific cluster was used to the calculations.


\begin{thebibliography}{10}
\expandafter\ifx\csname url\endcsname\relax
  \def\url#1{{\tt #1}}\fi
\expandafter\ifx\csname urlprefix\endcsname\relax\def\urlprefix{URL }\fi
\providecommand{\eprint}[2][]{\url{#2}}

\bibitem{shaf18}
Shaffer J~P, Rittenhouse S~T and Sadeghpour H~R 2018 {\em Nature
  Communications\/} {\bf 9} 1965

\bibitem{gree00}
Greene C~H, Dickinson A~S and Sadeghpour H~R 2000 {\em Phys. Rev. Lett.\/} {\bf
  85} 2458--2461

\bibitem{bend09}
Bendkowsky V, Butscher B, Nipper J, Shaffer J~P, L{\"o}w R and Pfau T 2009 {\em
  Nature\/} {\bf 458} 1005

\bibitem{li11}
Li W, Pohl T, Rost J~M, Rittenhouse S~T, Sadeghpour H~R, Nipper J, Butscher B,
  Balewski J~B, Bendkowsky V, L{\"o}w R and Pfau T 2011 {\em Science\/} {\bf
  334} 1110--1114 

\bibitem{tall12}
Tallant J, Rittenhouse S~T, Booth D, Sadeghpour H~R and Shaffer J~P 2012 {\em
  Phys. Rev. Lett.\/} {\bf 109} 173202

\bibitem{bell13}
Bellos M~A, Carollo R, Banerjee J, Eyler E~E, Gould P~L and Stwalley W~C 2013
  {\em Phys. Rev. Lett.\/} {\bf 111} 053001

\bibitem{ande14}
Anderson D~A, Miller S~A and Raithel G 2014 {\em Phys. Rev. Lett.\/} {\bf 112}
  163201

\bibitem{desa15}
DeSalvo B~J, Aman J~A, Dunning F~B, Killian T~C, Sadeghpour H~R, Yoshida S and
  Burgd\"orfer J 2015 {\em Phys. Rev. A\/} {\bf 92} 031403

\bibitem{sass15}
Sa\ss{}mannshausen H, Merkt F and Deiglmayr J 2015 {\em Phys. Rev. Lett.\/}
  {\bf 114} 133201

\bibitem{nied16}
Niederpr\"um T, Thomas O, Eichert T and Ott H 2016 {\em Phys. Rev. Lett.\/}
  {\bf 117} 123002

\bibitem{ferm34}
Fermi E 1934 {\em Nuovo Cimento\/} {\bf 11} 157

\bibitem{nara99}
Naraschewski M and Glauber R~J 1999 {\em Phys. Rev. A\/} {\bf 59} 4595--4607

\bibitem{whal19a}
Whalen J~D, Ding R, Kanungo S~K and Dunning F~B 2019 {\em Mol. Phys.\/}
  \urlprefix\url{https://doi.org/10.1080/00268976.2019.1575485}

\bibitem{whal19}
Whalen J~D, Kanungo S~K, Ding R, Wagner M, Schmidt R, Sadeghpour H~R, Yoshida
  S, Burgd\"orfer J, Dunning F~B and Killian T~C 2019 {\em Phys. Rev. A\/} {\bf
  100} 011402

\bibitem{dees09}
de~Escobar Y~N~M, Mickelson P~G, Yan M, DeSalvo B~J, Nagel S~B and Killian T~C
  2009 {\em Phys. Rev. Lett.\/} {\bf 103} 200402

\bibitem{stel14}
Stellmer S, Schreck F and Killian T~C 2014 Degenerate quantum gases of
  strontium {\em Annual Review of Cold Atoms and Molecules\/} vol~2 ed Madison
  K~W, Bongs K, Carr L~D, Rey A~M and Zhai H~ (World Scientific, Singapore)
  chap~1,  1--80

\bibitem{nage03}
Nagel S~B, Simien C~E, Laha S, Gupta P, Ashoka V~S and Killian T~C 2003 {\em
  Phys. Rev. A\/} {\bf 67}(1) 011401

\bibitem{thom18}
Thomas O, Lippe C, Eichert T and Ott H 2018 {\em J. Phys. B\/} {\bf 51} 155201

\bibitem{sous19}
Sous J, Sadeghpour H~R, Killian T~C, Demler E and Schmidt R 2019 Rydberg
  impurity in a fermi gas: Quantum statistics and rotational blockade
  arXiv:1907.07685v1

\bibitem{mazu17}
Mazurenko A, Chiu C~S, Ji G, Parsons M~F, Kan{\'a}sz-Nagy M, Schmidt R, Grusdt
  F, Demler E, Greif D and Greiner M 2017 {\em Nature\/} {\bf 545} 462

\bibitem{bakr10}
Bakr W~S, Peng A, Tai M~E, Ma R, Simon J, Gillen J~I, F{\"o}lling S, Pollet L
  and Greiner M 2010 {\em Science\/} {\bf 329} 547--550 

\bibitem{burt97}
Burt E~A, Ghrist R~W, Myatt C~J, Holland M~J, Cornell E~A and Wieman C~E 1997
  {\em Phys. Rev. Lett.\/} {\bf 79} 337--340

\bibitem{koeh06}
K\"ohler T, G\'oral K and Julienne P~S 2006 {\em Rev. Mod. Phys.\/} {\bf 78}
  1311--1361

\bibitem{jens04}
Jensen A~S, Riisager K, Fedorov D~V and Garrido E 2004 {\em Rev. Mod. Phys.\/}
  {\bf 76} 215--261

\bibitem{chin10}
Chin C, Grimm R, Julienne P and Tiesinga E 2010 {\em Rev. Mod. Phys.\/} {\bf
  82} 1225--1286

\end{thebibliography}
\providecommand{\newblock}{}

\end{document}